\documentclass[italian,english]{article}
\usepackage[T1]{fontenc}
\usepackage[latin1]{inputenc}
\usepackage{color}
\usepackage{graphicx}
\usepackage{amssymb}

\makeatletter

 \newcommand{\lyxaddress}[1]{
   \par {\raggedright #1 
   \vspace{1.4em}
   \noindent\par}
 }

\usepackage{babel}
\makeatother
\begin{document}

\title{\textbf{Massive gravitational waves from the $R^{2}$ theory of gravity:
production and response of interferometers }}

\author{\textbf{Christian Corda}}

\maketitle

\lyxaddress{\begin{center}INFN - Sezione di Pisa and Università di Pisa, Via
F. Buonarroti 2, I - 56127 PISA, Italy\end{center}}

\lyxaddress{\begin{center}\textit{E-mail address:} \textcolor{blue}{christian.corda@ego-gw.it} \end{center}}

\begin{abstract}
We show that from the \textbf{$R^{2}$} high order gravity theory
it is possible to produce, in the linearized approch, particles which
can be seen like massive modes of gravitational waves (GWs). The presence
of the mass generates a longitudinal force in addition of the transverse
one which is proper of the massless gravitational waves and the response
an interferometer to the effect is computed. This could be, in principle,
important to discriminate among the gravity theories. The presence
of the mass could also have important applications in cosmology because
the fact that gravitational waves can have mass could give a contribution
to the dark matter of the Universe.
\end{abstract}

\lyxaddress{PACS numbers: 04.80.Nn, 04.30.Nk, 04.50.+h}

\section{Introduction}

The data analysis of interferometric GWs detectors has recently started
(for the current status of GWs interferometers see \cite{key-1,key-2,key-3,key-4,key-5,key-6,key-7,key-8})
and the scientific community hopes in a first direct detection of
GWs in next years. The results of these detectors will have a fundamental
impact on astrophysics and gravitation physics. There will be lots
of experimental data to be analyzed, and theorists will be forced
to interact with lots of experiments and data analysts to extract
the physics from the data stream.

Detectors for GWs will be important for a better knowledge of the
Universe and also to confirm or ruling out the physical consistency
of General Relativity or of any other theory of gravitation \cite{key-9,key-10,key-11,key-12,key-13,key-14}.
This is because, in the context of Extended Theories of Gravity, some
differences between General Relativity and the others theories can
be pointed out starting by the linearized theory of gravity \cite{key-9,key-10,key-12,key-14}. 

In this paper the production and the potential detection with interferometers
of a hypotetical massive component of gravitational radiation which
arises from the \textbf{$R^{2}$} theory of gravity, which was the
first and simplest high order gravity theory proposed \cite{key-15},
is shown. 

In the second Section of this paper it is shown that a massive mode
of gravitational radiation arises from the high order action \cite{key-15}\begin{equation}
S=\int d^{4}x\sqrt{-g}(R+\alpha R^{2})+\mathcal{L}_{m}.\label{eq: high order 1}\end{equation}

Equation (\ref{eq: high order 1}) is a particular choice with respect
the well known canonical one of general relativity (the Einstein -
Hilbert action \cite{key-16,key-17}) which is 

\begin{equation}
S=\int d^{4}x\sqrt{-g}R+\mathcal{L}_{m},\label{eq: EH}\end{equation}

where $R$ is the Ricci scalar curvature. We empahsize that the presence
of the mass could also have important applications in cosmology because
the fact that gravitational waves can have mass could give a contribution
to the dark matter of the Universe. We also recall that an alternative
way to resolve the dark matter and dark energy problems using high
order gravity is shown in ref. \cite{key-18}.

In Section three it is shown that the massive component generates
a longitudinal force in addition of the transverse one which is proper
of the massless case. 

After this, in Section four, the potential interferometric detection
of this longitudinal component is analyzed and the response of an
interferometer is computed. This could be, in principle, important
to discriminate among several gravity theories which are today considered.

\section{The production of a massive mode of gravitational radiation in the
$R^{2}$ theory of gravity }

If the gravitational Lagrangian is nonlinear in the curvature invariants
the Einstein field equations has an order higher than second \cite{key-9,key-12,key-13}.
For this reason such theories are often called higher-order gravitational
theories. This is exactly the case of the action (\ref{eq: high order 1}).

By varying this action with respect to $g_{\mu\nu}$ (see refs. \cite{key-12,key-13}
for a parallel computation) the field equations are obtained (note
that in this paper we work with $G=1$, $c=1$ and $\hbar=1$):

\begin{equation}
\begin{array}{c}
G_{\mu\nu}=\frac{-4\pi\tilde{G}}{2\alpha R+1}\{+T_{\mu\nu}^{(m)}-\frac{1}{2}g_{\mu\nu}\alpha R^{2}+\\
\\+2\alpha R_{;\mu;\nu}-2\alpha g_{\mu\nu}\square R\}\end{array}\label{eq: einstein-general}\end{equation}

with associed a Klein - Gordon equation for the Ricci curvature scalar 

\begin{equation}
\square R=m^{2}(R+8\pi\tilde{G}T),\label{eq: KG}\end{equation}

where $\square$ is the d' Alembertian operator and the mass $m$
has been introduced for dimensional motivations: $m^{2}\equiv-\frac{1}{6\alpha}$,
thus $\alpha$ has to be negative \cite{key-15}.

In the above equations $T_{\mu\nu}^{(m)}$ is the ordinary stress-energy
tensor of the matter and $\tilde{G}$ is a dimensional, strictly positive,
constant \cite{key-9,key-12,key-13}. The Newton constant is replaced
by the effective coupling

\begin{equation}
G_{eff}=-\frac{1}{2(2\alpha R+1)},\label{eq: newton eff}\end{equation}

which is different from $G$. General relativity is obtained when
$\alpha=0.$

To study gravitational waves the linearized theory in vacuum ($T_{\mu\nu}^{(m)}=0$)
has to be analyzed, with a little perturbation of the background,
which is assumed given by the Minkowskian background. In this case
the Ricci scalar is assumed slowly varying near zero: $R\simeq0+\delta R\equiv h_{R}.$ 

Putting

\begin{equation}
g_{\mu\nu}=\eta_{\mu\nu}+h_{\mu\nu}\label{eq: linearizza}\end{equation}

to first order in $h_{\mu\nu}$ , calling $\widetilde{R}_{\mu\nu\rho\sigma}$
, $\widetilde{R}_{\mu\nu}$ and $\widetilde{R}$ the linearized quantity
which correspond to $R_{\mu\nu\rho\sigma}$ , $R_{\mu\nu}$ and $R$,
the linearized field equations are obtained \cite{key-12,key-13,key-16,key-17}:

\begin{equation}
\begin{array}{c}
\widetilde{R}_{\mu\nu}-\frac{\widetilde{R}}{2}\eta_{\mu\nu}=\partial_{\mu}\partial_{\nu}\widetilde{R}+\eta_{\mu\nu}\square h_{R}\\
\\{}\square h_{R}=m^{2}h_{R}.\end{array}\label{eq: linearizzate1}\end{equation}

$\widetilde{R}_{\mu\nu\rho\sigma}$ and eqs. (\ref{eq: linearizzate1})
are invariants for gauge transformations \cite{key-12,key-13}

\begin{equation}
\begin{array}{c}
h_{\mu\nu}\rightarrow h'_{\mu\nu}=h_{\mu\nu}-\partial_{(\mu}\epsilon_{\nu)}\\
\\h_{R}\rightarrow h_{R}'=h_{R};\end{array}\label{eq: gauge}\end{equation}

then 

\begin{equation}
\bar{h}_{\mu\nu}\equiv h_{\mu\nu}-\frac{h}{2}\eta_{\mu\nu}+\eta_{\mu\nu}h_{R}\label{eq: ridefiniz}\end{equation}

can be defined, and, considering the transform for the parameter $\epsilon^{\mu}$

\begin{equation}
\square\epsilon_{\nu}=\partial^{\mu}\bar{h}_{\mu\nu},\label{eq:lorentziana}\end{equation}
 a gauge parallel to the Lorenz one of electromagnetic waves can be
choosen:

\begin{equation}
\partial^{\mu}\bar{h}_{\mu\nu}=0.\label{eq: cond lorentz}\end{equation}

In this way field equations read like

\begin{equation}
\square\bar{h}_{\mu\nu}=0\label{eq: onda T}\end{equation}

\begin{equation}
\square h_{R}=m^{2}h_{R}\label{eq: onda S}\end{equation}

Solutions of eqs. (\ref{eq: onda T}) and (\ref{eq: onda S}) are
plan waves:

\begin{equation}
\bar{h}_{\mu\nu}=A_{\mu\nu}(\overrightarrow{p})\exp(ip^{\alpha}x_{\alpha})+c.c.\label{eq: sol T}\end{equation}

\begin{equation}
h_{R}=a(\overrightarrow{p})\exp(iq^{\alpha}x_{\alpha})+c.c.\label{eq: sol S}\end{equation}

where

\begin{equation}
\begin{array}{ccc}
k^{\alpha}\equiv(\omega,\overrightarrow{p}) &  & \omega=p\equiv|\overrightarrow{p}|\\
\\q^{\alpha}\equiv(\omega_{m},\overrightarrow{p}) &  & \omega_{m}=\sqrt{m^{2}+p^{2}}.\end{array}\label{eq: k e q}\end{equation}

In eqs. (\ref{eq: onda T}) and (\ref{eq: sol T}) the equation and
the solution for the waves like in standard general relativity \cite{key-16,key-17}
have been obtained, but eqs. (\ref{eq: onda S}) and (\ref{eq: sol S})
are respectively the equation and the solution for the massive mode
(see also \cite{key-12,key-13}) arising from the Starobinsky's high
order gravity theory.

The fact that the dispersion law for the modes of the massive field
$h_{R}$ is not linear has to be emphatized. The velocity of every
tensorial mode $\bar{h}_{\mu\nu}$ is the light speed $c$, but the
dispersion law (the second of eq. (\ref{eq: k e q})) for the modes
of $h_{R}$ is that of a massive field which can be discussed like
a wave-packet \cite{key-12,key-13}. Also, the group-velocity of a
wave-packet of $h_{R}$ centered in $\overrightarrow{p}$ is 

\begin{equation}
\overrightarrow{v_{G}}=\frac{\overrightarrow{p}}{\omega},\label{eq: velocita' di gruppo}\end{equation}

which is exactly the velocity of a massive particle with mass $m$
and momentum $\overrightarrow{p}$.

From the second of eqs. (\ref{eq: k e q}) and eq. (\ref{eq: velocita' di gruppo})
it is simple to obtain:

\begin{equation}
v_{G}=\frac{\sqrt{\omega^{2}-m^{2}}}{\omega}.\label{eq: velocita' di gruppo 2}\end{equation}

Then, wanting a constant speed of our wave-packet, it has to be \cite{key-12,key-13}

\begin{equation}
m=\sqrt{(1-v_{G}^{2})}\omega.\label{eq: relazione massa-frequenza}\end{equation}

The relation (\ref{eq: relazione massa-frequenza}) is shown in fig.
1 for a value $v_{G}=0.9$.

\begin{figure}
\includegraphics{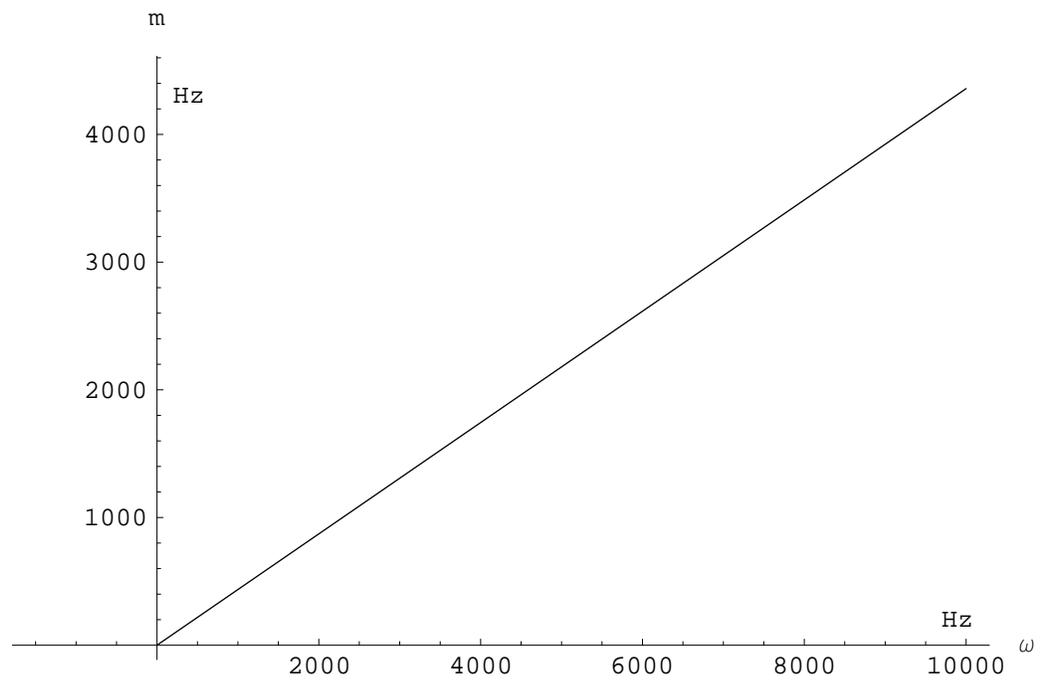}

\caption{the mass-frequency relation for a massive graviiatonal wave arising
from the \textbf{$R^{2}$} high order gravity theory and propagating
with a speed of $0.9c$ : for the mass it is $1Hz=10^{-15}eV$}
\end{figure}

Now the analisys can remain in the Lorenz gauge with trasformations
of the type $\square\epsilon_{\nu}=0$; this gauge gives a condition
of transversality for the tensorial part of the field: $k^{\mu}A_{\mu\nu}=0$,
but does not give the transversality for the total field $h_{\mu\nu}$.
From eq. (\ref{eq: ridefiniz}) it is

\begin{equation}
h_{\mu\nu}=\bar{h}_{\mu\nu}-\frac{\bar{h}}{2}\eta_{\mu\nu}+\eta_{\mu\nu}h_{R}.\label{eq: ridefiniz 2}\end{equation}

At this point, if being in the massless case \cite{key-17}, it could
been put

\begin{equation}
\begin{array}{c}
\square\epsilon^{\mu}=0\\
\\\partial_{\mu}\epsilon^{\mu}=-\frac{\bar{h}}{2}+h_{R},\end{array}\label{eq: gauge2}\end{equation}

which gives the total transversality of the field. But in the massive
case this is impossible. In fact, applying the Dalembertian operator
to the second of eqs. (\ref{eq: gauge2}) and using the field equations
(\ref{eq: onda T}) and (\ref{eq: onda S}) it results

\begin{equation}
\square\epsilon^{\mu}=m^{2}h_{R},\label{eq: contrasto}\end{equation}

which is in contrast with the first of eqs. (\ref{eq: gauge2}). In
the same way it is possible to show that it does not exist any linear
relation between the field $\bar{h}_{\mu\nu}$ and $h_{R}$. Thus
a gauge in wich $h_{\mu\nu}$ is purely spatial cannot be chosen (i.e.
it cannot be put $h_{\mu0}=0,$ see eq. (\ref{eq: ridefiniz 2}))
. But the traceless condition to the field $\bar{h}_{\mu\nu}$ can
be put:

\begin{equation}
\begin{array}{c}
\square\epsilon^{\mu}=0\\
\\\partial_{\mu}\epsilon^{\mu}=-\frac{\bar{h}}{2}.\end{array}\label{eq: gauge traceless}\end{equation}

These equations imply

\begin{equation}
\partial^{\mu}\bar{h}_{\mu\nu}=0.\label{eq: vincolo}\end{equation}

To save the conditions $\partial_{\mu}\bar{h}^{\mu\nu}$ and $\bar{h}=0$
transformations like

\begin{equation}
\begin{array}{c}
\square\epsilon^{\mu}=0\\
\\\partial_{\mu}\epsilon^{\mu}=0\end{array}\label{eq: gauge 3}\end{equation}

can be used and, taking $\overrightarrow{p}$ in the $z$ direction,
a gauge in which only $A_{11}$, $A_{22}$, and $A_{12}=A_{21}$ are
different to zero can be chosen. The condition $\bar{h}=0$ gives
$A_{11}=-A_{22}$. Now, putting these equations in eq. (\ref{eq: ridefiniz 2})
it results

\begin{equation}
h_{\mu\nu}(t,z)=A^{+}(t-z)e_{\mu\nu}^{(+)}+A^{\times}(t-z)e_{\mu\nu}^{(\times)}+h_{R}(t-v_{G}z)\eta_{\mu\nu}.\label{eq: perturbazione totale}\end{equation}

The term $A^{+}(t-z)e_{\mu\nu}^{(+)}+A^{\times}(t-z)e_{\mu\nu}^{(\times)}$
describes the two standard polarizations of gravitational waves which
arise from General Relativity, while the term $h_{R}(t-v_{G}z)\eta_{\mu\nu}$
is the massive polarization arising from the \textbf{$R^{2}$} theory.

\section{The presence of a longitudinal force}

The analysis of the two standard polarization is well known in the
literature \cite{key-16,key-17}. For a the pure polarization arising
by the \textbf{$R^{2}$} theory eq. (\ref{eq: perturbazione totale})
can be rewritten as

\begin{equation}
h_{\mu\nu}(t-v_{G}z)=h_{R}(t-v_{G}z)\eta_{\mu\nu}\label{eq: perturbazione scalare}\end{equation}
and the corrispondent line element is the conformally flat one

\begin{equation}
ds^{2}=[1+h_{R}(t-v_{G}z)](-dt^{2}+dz^{2}+dx^{2}+dy^{2}).\label{eq: metrica puramente scalare}\end{equation}
But, in a laboratory environment on Earth, the coordinate system in
which the space-time is locally flat is typically used and the distance
between any two points is given simply by the difference in their
coordinates in the sense of Newtonian physics \cite{key-12,key-13,key-16,key-17}.
This frame is the proper reference frame of a local observer, located
for example in the position of the beam splitter of an interferometer.
In this frame gravitational waves manifest themself by exerting tidal
forces on the masses (the mirror and the beam-splitter in the case
of an interferometer). A detailed analysis of the frame of the local
observer is given in ref. \cite{key-17}, sect. 13.6. Here only the
more important features of this coordinate system are recalled:

the time coordinate $x_{0}$ is the proper time of the observer O;

spatial axes are centered in O;

in the special case of zero acceleration and zero rotation the spatial
coordinates $x_{j}$ are the proper distances along the axes and the
frame of the local observer reduces to a local Lorentz frame: in this
case the line element reads \cite{key-17}

\begin{equation}
ds^{2}=-(dx^{0})^{2}+\delta_{ij}dx^{i}dx^{j}+O(|x^{j}|^{2})dx^{\alpha}dx^{\beta}.\label{eq: metrica local lorentz}\end{equation}

The effect of the gravitational wave on test masses is described by
the equation

\begin{equation}
\ddot{x^{i}}=-\widetilde{R}_{0k0}^{i}x^{k},\label{eq: deviazione geodetiche}\end{equation}
which is the equation for geodesic deviation in this frame.

Thus, to study the effect of the massive gravitational wave on test
masses, $\widetilde{R}_{0k0}^{i}$ has to be computed in the proper
reference frame of the local observer. But, because the linearized
Riemann tensor $\widetilde{R}_{\mu\nu\rho\sigma}$ is invariant under
gauge transformations \cite{key-12,key-13,key-17}, it can be directly
computed from eq. (\ref{eq: perturbazione scalare}). 

From \cite{key-17} it is:

\begin{equation}
\widetilde{R}_{\mu\nu\rho\sigma}=\frac{1}{2}\{\partial_{\mu}\partial_{\beta}h_{\alpha\nu}+\partial_{\nu}\partial_{\alpha}h_{\mu\beta}-\partial_{\alpha}\partial_{\beta}h_{\mu\nu}-\partial_{\mu}\partial_{\nu}h_{\alpha\beta}\},\label{eq: riemann lineare}\end{equation}

that, in the case eq. (\ref{eq: perturbazione scalare}), begins

\begin{equation}
\widetilde{R}_{0\gamma0}^{\alpha}=\frac{1}{2}\{\partial^{\alpha}\partial_{0}h_{R}\eta_{0\gamma}+\partial_{0}\partial_{\gamma}h_{R}\delta_{0}^{\alpha}-\partial^{\alpha}\partial_{\gamma}h_{R}\eta_{00}-\partial_{0}\partial_{0}h_{R}\delta_{\gamma}^{\alpha}\};\label{eq: riemann lin scalare}\end{equation}

the different elements are (only the non zero ones will be written):

\begin{equation}
\partial^{\alpha}\partial_{0}h_{R}\eta_{0\gamma}=\left\{ \begin{array}{ccc}
\partial_{t}^{2}h_{R} & for & \alpha=\gamma=0\\
\\-\partial_{z}\partial_{t}h_{R} & for & \alpha=3;\gamma=0\end{array}\right\} \label{eq: calcoli}\end{equation}

\begin{equation}
\partial_{0}\partial_{\gamma}h_{R}\delta_{0}^{\alpha}=\left\{ \begin{array}{ccc}
\partial_{t}^{2}h_{R} & for & \alpha=\gamma=0\\
\\\partial_{t}\partial_{z}h_{R} & for & \alpha=0;\gamma=3\end{array}\right\} \label{eq: calcoli2}\end{equation}

\begin{equation}
-\partial^{\alpha}\partial_{\gamma}h_{R}\eta_{00}=\partial^{\alpha}\partial_{\gamma}h_{R}=\left\{ \begin{array}{ccc}
-\partial_{t}^{2}h_{R} & for & \alpha=\gamma=0\\
\\\partial_{z}^{2}h_{R} & for & \alpha=\gamma=3\\
\\-\partial_{t}\partial_{z}h_{R} & for & \alpha=0;\gamma=3\\
\\\partial_{z}\partial_{t}h_{R} & for & \alpha=3;\gamma=0\end{array}\right\} \label{eq: calcoli3}\end{equation}

\begin{equation}
-\partial_{0}\partial_{0}h_{R}\delta_{\gamma}^{\alpha}=\begin{array}{ccc}
-\partial_{z}^{2}h_{R} & for & \alpha=\gamma\end{array}.\label{eq: calcoli4}\end{equation}

Now, putting these results in eq. (\ref{eq: riemann lin scalare})
it results:

\begin{equation}
\begin{array}{c}
\widetilde{R}_{010}^{1}=-\frac{1}{2}\ddot{h}_{R}\\
\\\widetilde{R}_{010}^{2}=-\frac{1}{2}\ddot{h}_{R}\\
\\\widetilde{R}_{030}^{3}=\frac{1}{2}\square h_{R}.\end{array}\label{eq: componenti riemann}\end{equation}

But, putting the field equation (\ref{eq: onda S}) in the third of
eqs. (\ref{eq: componenti riemann}) it is

\begin{equation}
\widetilde{R}_{030}^{3}=\frac{1}{2}m^{2}h_{R},\label{eq: terza riemann}\end{equation}

which shows that the field is not transversal. 

Infact, using eq. (\ref{eq: deviazione geodetiche}) it results

\begin{equation}
\ddot{x}=\frac{1}{2}\ddot{h}_{R}x,\label{eq: accelerazione mareale lungo x}\end{equation}

\begin{equation}
\ddot{y}=\frac{1}{2}\ddot{h}_{R}y\label{eq: accelerazione mareale lungo y}\end{equation}

and 

\begin{equation}
\ddot{z}=-\frac{1}{2}m^{2}h_{R}(t-v_{G}z)z.\label{eq: accelerazione mareale lungo z}\end{equation}

Then the effect of the mass is the generation of a \textit{longitudinal}
force (in addition to the transverse one). Note that in the limit
$m\rightarrow0$ the longitudinal force vanishes.

\section{The interferometer's response to the longitudinal component}

Before starting the analysis it has to be discussed if there are fenomenogical
limitations to the mass of the wave \cite{key-12,key-13}. Treating
$h_{R}$ like a classical wave, that acts coherently with the interferometer,
it has to be $m\ll1/L$ , where $L=3$ kilometers in the case of Virgo
and $L=4$ kilometers in the case of LIGO. Thus it has to be approximately
$m<10^{-9}eV$. However there is a stronger limitation coming from
the fact that the massive wave needs a frequency which falls in the
frequency-range for earth based gravitational antennas that is the
interval $10Hz\leq f\leq10KHz$ \cite{key-1,key-2,key-3,key-4,key-5,key-6,key-7,key-8}.
For a massive gravitational wave, from the second of eqs. (\ref{eq: k e q})
it is:

\begin{equation}
2\pi f=\omega=\sqrt{m^{2}+p^{2}},\label{eq: frequenza-massa}\end{equation}

were $p$ is the momentum \cite{key-13}. Thus it needs

\begin{equation}
0eV\leq m\leq10^{-11}eV.\label{eq: range di massa}\end{equation}

For these light particles their effect can be still discussed as a
coherent gravitational wave. For the discussion of this longitudinal
effect we start directly from the gauge (\ref{eq: metrica puramente scalare}). 

Eq. (\ref{eq: metrica puramente scalare}) can be rewritten as

\begin{equation}
(\frac{dt}{d\tau})^{2}-(\frac{dx}{d\tau})^{2}-(\frac{dy}{d\tau})^{2}-(\frac{dz}{d\tau})^{2}=\frac{1}{(1+h_{R})},\label{eq: Sh2}\end{equation}

where $\tau$ is the proper time of the test masses.

From eqs. (\ref{eq: metrica puramente scalare}) and (\ref{eq: Sh2})
the geodesic equations of motion for test masses (i.e. the beam-splitter
and the mirrors of the interferometer), can be obtained\begin{equation}
\begin{array}{ccc}
\frac{d^{2}x}{d\tau^{2}} & = & 0\\
\\\frac{d^{2}y}{d\tau^{2}} & = & 0\\
\\\frac{d^{2}t}{d\tau^{2}} & = & \frac{1}{2}\frac{\partial_{t}(1+h_{R})}{(1+h_{R})^{2}}\\
\\\frac{d^{2}z}{d\tau^{2}} & = & -\frac{1}{2}\frac{\partial_{z}(1+h_{R})}{(1+h_{R})^{2}}.\end{array}\label{eq: geodetiche Corda}\end{equation}

The first and the second of eqs. (\ref{eq: geodetiche Corda}) can
be immediately integrated obtaining

\begin{equation}
\frac{dx}{d\tau}=C_{1}=const.\label{eq: integrazione x}\end{equation}

\begin{equation}
\frac{dy}{d\tau}=C_{2}=const.\label{eq: integrazione x}\end{equation}

In this way eq. (\ref{eq: Sh2}) becomes\begin{equation}
(\frac{dt}{d\tau})^{2}-(\frac{dz}{d\tau})^{2}=\frac{1}{(1+h_{R})}.\label{eq: Ch3}\end{equation}

If we assume that test masses are at rest initially we get $C_{1}=C_{2}=0$.
Thus we see that, even if the GW arrives at test masses, we do not
have motion of test masses within the $x-y$ plane in this gauge.
We could understand this directly from eq. (\ref{eq: metrica puramente scalare})
because the absence of the $x$ and of the $y$ dependences in the
metric implies that test masses momentum in these directions (i.e.
$C_{1}$ and $C_{2}$ respectively) is conserved. This results, for
example, from the fact that in this case the $x$ and $y$ coordinates
do not esplicitly enter in the Hamilton-Jacobi equation for a test
mass in a gravitational field \cite{key-16}. 

Now we will see that, in presence of the GW, we have motion of test
masses in the $z$ direction which is the direction of the propagating
wave. An analysis of eqs. (\ref{eq: geodetiche Corda}) shows that,
to simplify equations, we can introduce the retarded and advanced
time coordinates ($u,v$):

\begin{equation}
\begin{array}{c}
u=t-v_{G}z\\
\\v=t+v_{G}z.\end{array}\label{eq: ret-adv}\end{equation}

From the third and the fourth of eqs. (\ref{eq: geodetiche Corda})
we have

\begin{equation}
\frac{d}{d\tau}\frac{du}{d\tau}=\frac{\partial_{v}[1+h_{R}(u)]}{(1+h_{R}(u))^{2}}=0.\label{eq: t-z t+z}\end{equation}

This equation can be integrated obtaining

\begin{equation}
\frac{du}{d\tau}=\alpha,\label{eq: t-z}\end{equation}

where $\alpha$ is an integration constant. From eqs. (\ref{eq: Ch3})
and (\ref{eq: t-z}), we also get

\begin{equation}
\frac{dv}{d\tau}=\frac{\beta}{1+h_{R}}\label{eq: t+z}\end{equation}

where $\beta\equiv\frac{1}{\alpha}$, and

\begin{equation}
\tau=\beta u+\gamma,\label{eq: tau}\end{equation}

where the integration constant $\gamma$ correspondes simply to the
retarded time coordinate translation $u$. Thus, without loss of generality,
we can put it equal to zero. Now let us see what is the meaning of
the other integration constant $\beta.$ We can write the equation
for $z$ from eqs. (\ref{eq: t-z}) and (\ref{eq: t+z}):

\begin{equation}
\frac{dz}{d\tau}=\frac{1}{2\beta}(\frac{\beta^{2}}{1+h_{R}}-1).\label{eq: z}\end{equation}

When it is $h_{R}=0$ (i.e. before the GW arrives at the test masses)
eq. (\ref{eq: z}) becomes\begin{equation}
\frac{dz}{d\tau}=\frac{1}{2\beta}(\beta^{2}-1).\label{eq: z ad h nullo}\end{equation}

But this is exactly the initial velocity of the test mass, so we have
to choose $\beta=1$ because we suppose that test masses are at rest
initially. This also imply $\alpha=1$.

To find the motion of a test mass in the $z$ direction we see that
from eq. (\ref{eq: tau}) we have $d\tau=du$, while from eq. (\ref{eq: t+z})
we have $dv=\frac{d\tau}{1+h_{R}}$. Because it is $v_{G}z=\frac{v-u}{2}$
we obtain

\begin{equation}
dz=\frac{1}{2v_{G}}(\frac{d\tau}{1+h_{R}}-du),\label{eq: dz}\end{equation}

which can be integrated as

\begin{equation}
\begin{array}{c}
z=z_{0}+\frac{1}{2v_{G}}\int(\frac{du}{1+h_{R}}-du)=\\
\\=z_{0}-\frac{1}{2v_{G}}\int_{-\infty}^{t-v_{G}z}\frac{h_{R}(u)}{1+h_{R}(u)}du,\end{array}\label{eq: moto lungo z}\end{equation}

where $z_{0}$ is the initial position of the test mass. Now the displacement
of the test mass in the $z$ direction can be written as

\begin{equation}
\begin{array}{c}
\Delta z=z-z_{0}=-\frac{1}{2v_{G}}\int_{-\infty}^{t-v_{G}z_{0}-v_{G}\Delta z}\frac{h_{R}(u)}{1+h_{R}(u)}du\\
\\\simeq-\frac{1}{2v_{G}}\int_{-\infty}^{t-v_{G}z_{0}}\frac{h_{R}(u)}{1+h_{R}(u)}du.\end{array}\label{eq: spostamento lungo z}\end{equation}
We can also rewrite our results in function of the time coordinate
$t$:

\begin{equation}
\begin{array}{ccc}
x(t) & = & x_{0}\\
\\y(t) & = & y_{0}\\
\\z(t) & = & z_{0}-\frac{1}{2v_{G}}\int_{-\infty}^{t-v_{G}z_{0}}\frac{h_{R}(u)}{1+h_{R}(u)}d(u)\\
\\\tau(t) & = & t-v_{G}z(t),\end{array}\label{eq: moto gauge Corda}\end{equation}

Calling $l$ and $L+l$ the unperturbed positions of the beam-splitter
and of the mirror and using the third of eqs. (\ref{eq: moto gauge Corda})
the varying position of the beam-splitter and of the mirror are given
by

\begin{equation}
\begin{array}{c}
z_{BS}(t)=l-\frac{1}{2v_{G}}\int_{-\infty}^{t-v_{G}l}\frac{h_{R}(u)}{1+h_{R}(u)}d(u)\\
\\z_{M}(t)=L+l-\frac{1}{2v_{G}}\int_{-\infty}^{t-v_{G}(L+l)}\frac{h_{R}(u)}{1+h_{R}(u)}d(u)\end{array}\label{eq: posizioni}\end{equation}

But we are interested in variations in the proper distance (time)
of test masses, thus, in correspondence of eqs. (\ref{eq: posizioni}),
using the fourth of eqs. (\ref{eq: moto gauge Corda}) we get\begin{equation}
\begin{array}{c}
\tau_{BS}(t)=t-v_{G}l-\frac{1}{2}\int_{-\infty}^{t-v_{G}l}\frac{h_{R}(u)}{1+h_{R}(u)}d(u)\\
\\\tau_{M}(t)=t-v_{G}L-v_{G}l-\frac{1}{2}\int_{-\infty}^{t-v_{G}(L+l)}\frac{h_{R}(u)}{1+h_{R}(u)}d(u).\end{array}\label{eq: posizioni 2}\end{equation}

Then the total variation of the proper time is given by

\begin{equation}
\bigtriangleup\tau(t)=\tau_{M}(t)-\tau_{BS}(t)=v_{G}L-\frac{1}{2}\int_{t-v_{G}l}^{t-v_{G}(L+l)}\frac{h_{R}(u)}{1+h_{R}(u)}d(u).\label{eq: time}\end{equation}

In this way, recalling that in the used units the unperturbed proper
distance (time)is $T=L$, the difference between the total variation
of the proper time in presence and the total variation of the proper
time in absence of the GW is \begin{equation}
\delta\tau(t)\equiv\bigtriangleup\tau(t)-L=-L(v_{G}+1)-\frac{1}{2}\int_{t-v_{G}l}^{t-v_{G}(L+l)}\frac{h_{R}(u)}{1+h_{R}(u)}d(u).\label{eq: time variation}\end{equation}

This quantity can be computed in the frequency domain, defining the
Fourier transform of $h_{R}$ as \begin{equation}
\widetilde{h}_{R}(\omega)=\int_{-\infty}^{\infty}dt\textrm{ }h_{R}(t)\exp(i\omega t).\label{eq: trasformata di fourier}\end{equation}

and using the translation and derivation Fourier theorems, obtaining\begin{equation}
\begin{array}{c}
\delta\widetilde{\tau}(\omega)=L(1-v_{G}^{2})\exp[i\omega L(1+v_{G})]+\frac{L}{2\omega L(v_{G}^{2}-1)^{2}}\\
\\{}[\exp[2i\omega L](v_{G}+1)^{3}(-2i+\omega L(v_{G}-1)+2L\exp[i\omega L(1+v_{G})]\\
\\(6iv_{G}+2iv_{G}^{3}-\omega L+\omega Lv_{G}^{4})+L(v_{G}+1)^{3}(-2i+\omega L(v_{G}+1))]\widetilde{h}_{R}.\end{array}\label{eq: segnale totale lungo z}\end{equation}

A {}``signal'' can be also defined:

\begin{equation}
\begin{array}{c}
\widetilde{S}(\omega)\equiv\frac{\delta\widetilde{\tau}(\omega)}{L}=(1-v_{G}^{2})\exp[i\omega L(1+v_{G})]+\frac{1}{2\omega L(v_{G}^{2}-1)^{2}}\\
\\{}[\exp[2i\omega L](v_{G}+1)^{3}(-2i+\omega L(v_{G}-1)+2\exp[i\omega L(1+v_{G})]\\
\\(6iv_{G}+2iv_{G}^{3}-\omega L+\omega Lv_{G}^{4})+(v_{G}+1)^{3}(-2i+\omega L(v_{G}+1))]\widetilde{h}_{R}.\end{array}\label{eq: sig}\end{equation}

Then the function \begin{equation}
\begin{array}{c}
\Upsilon_{l}(\omega)\equiv(1-v_{G}^{2})\exp[i\omega L(1+v_{G})]+\frac{1}{2\omega L(v_{G}^{2}-1)^{2}}\\
\\{}[\exp[2i\omega L](v_{G}+1)^{3}(-2i+\omega L(v_{G}-1)+2\exp[i\omega L(1+v_{G})]\\
\\(6iv_{G}+2iv_{G}^{3}-\omega L+\omega Lv_{G}^{4})+(v_{G}+1)^{3}(-2i+\omega L(v_{G}+1))],\end{array}\label{eq: risposta totale lungo z due}\end{equation}

is the response function of an arm of our interferometer located in
the $z$-axis, due to the longitudinal component of the massive gravitational
wave arising from the \textbf{$R^{2}$} high order gravity theory
and propagating in the same direction of the axis.

For $v_{G}\rightarrow1$ it is $\Upsilon_{l}(\omega)\rightarrow0$. 

In figures 2, 3 and 4 are shown the response functions (\ref{eq: risposta totale lungo z due})
for an arm of the Virgo interferometer ($L=3Km$) for $v_{G}=0.1$
(non-relativistic case), $v_{G}=0.9$ (relativistic case) and $v_{G}=0.999$
(ultra-relativistic case). We see that in the non-relativistic case
the signal is stronger as it could be expected (for $m\rightarrow0$
we expect$\Upsilon_{l}(\omega)\rightarrow0$). In figures 5, 6, and
7 the same response functions are shown for the Ligo interferometer
($L=4Km$).

\begin{figure}
\includegraphics{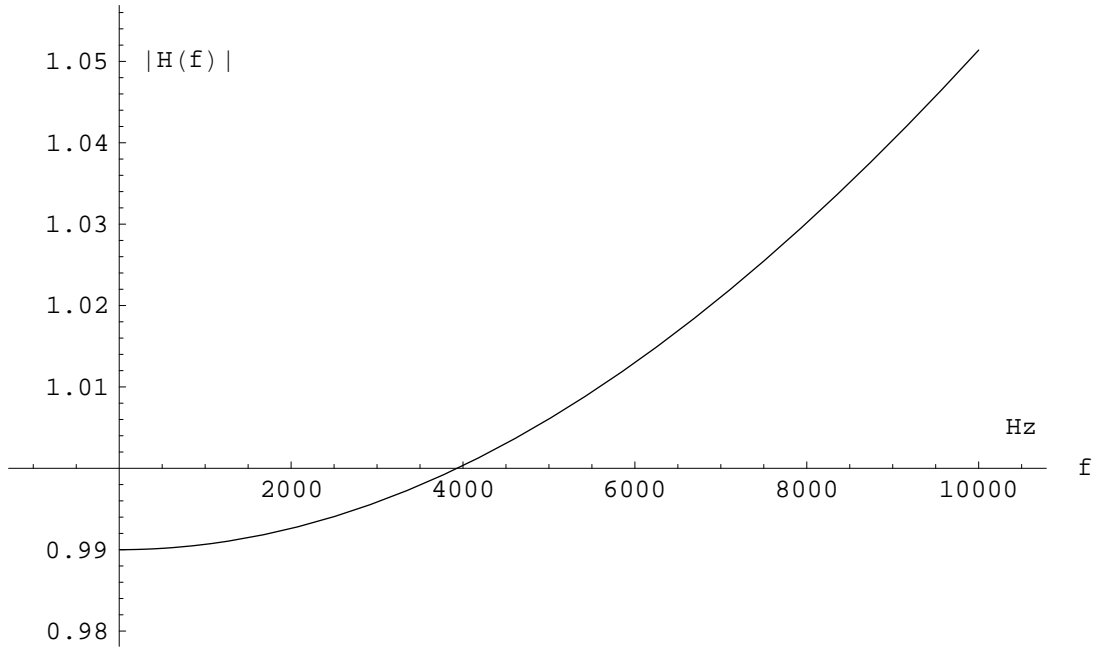}

\caption{the absolute value of the longitudinal response function (\ref{eq: segnale totale lungo z})
of the Virgo interferometer ($L=3Km$) to a GW arising from the \textbf{$R^{2}$}
high order gravity theory and propagating with a speed of $0.1c$
(non relativistic case). }
\end{figure}
\begin{figure}
\includegraphics{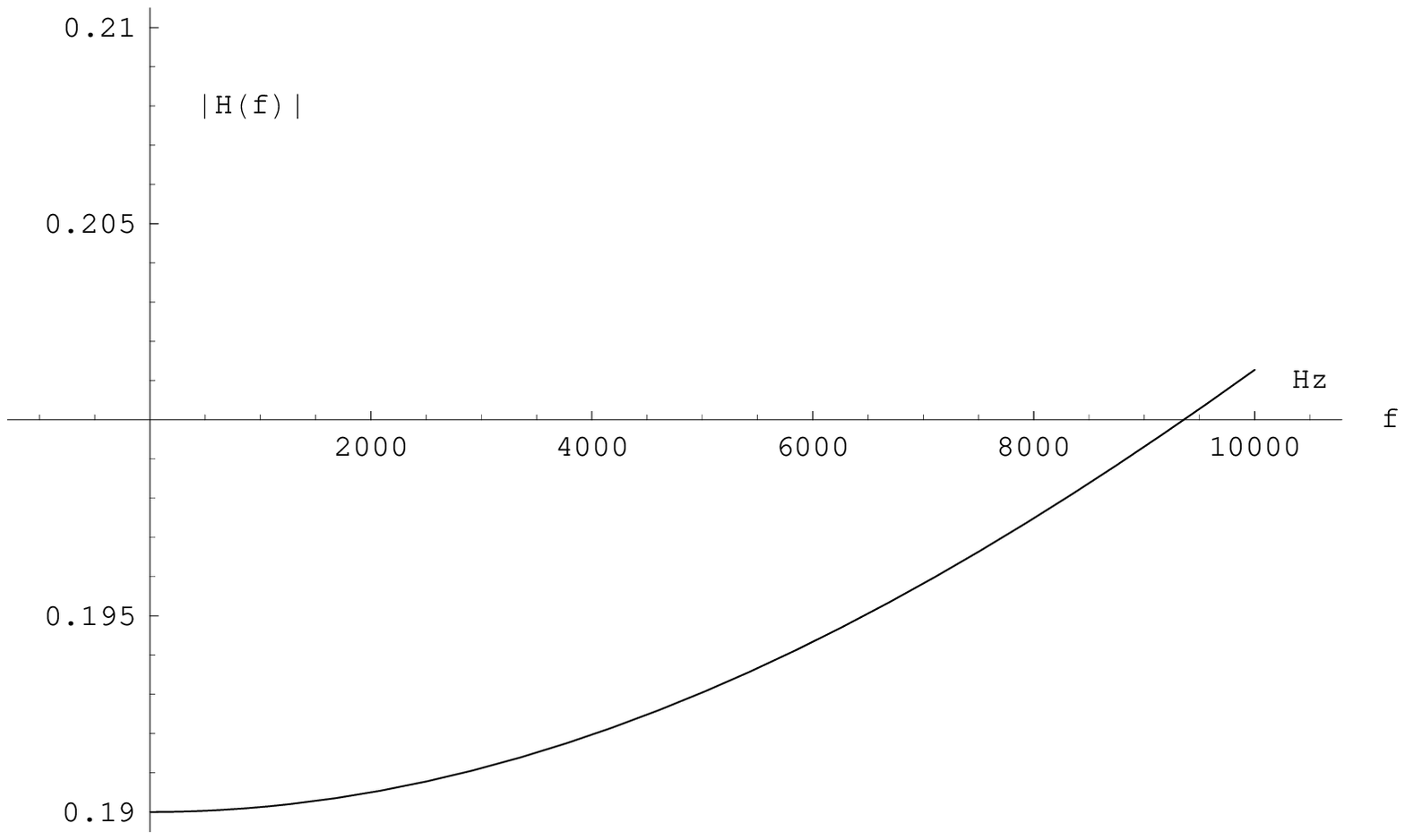}

\caption{the absolute value of the longitudinal response function (\ref{eq: segnale totale lungo z})
of the Virgo interferometer ($L=3Km$) to a GW arising from the \textbf{$R^{2}$}
high order gravity theory and propagating with a speed of $0.9$ (relativistic
case). }
\end{figure}
\begin{figure}
\includegraphics{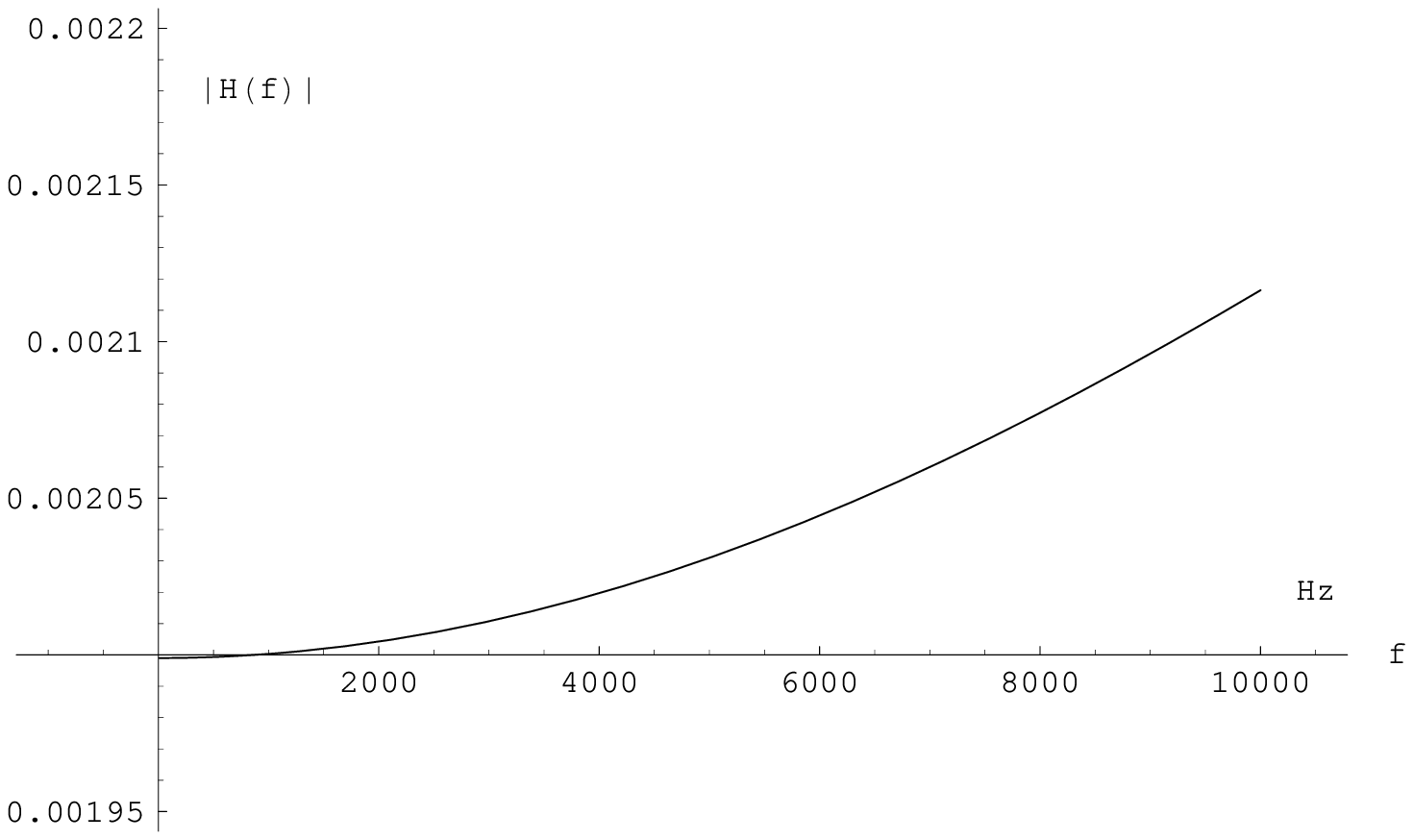}

\caption{the absolute value of the longitudinal response function (\ref{eq: segnale totale lungo z})
of the Virgo interferometer ($L=3Km$) to a GW arising from the \textbf{$R^{2}$}
high order gravity theory and propagating with a speed of $0.999$
(ultra relativistic case). }
\end{figure}
\begin{figure}
\includegraphics{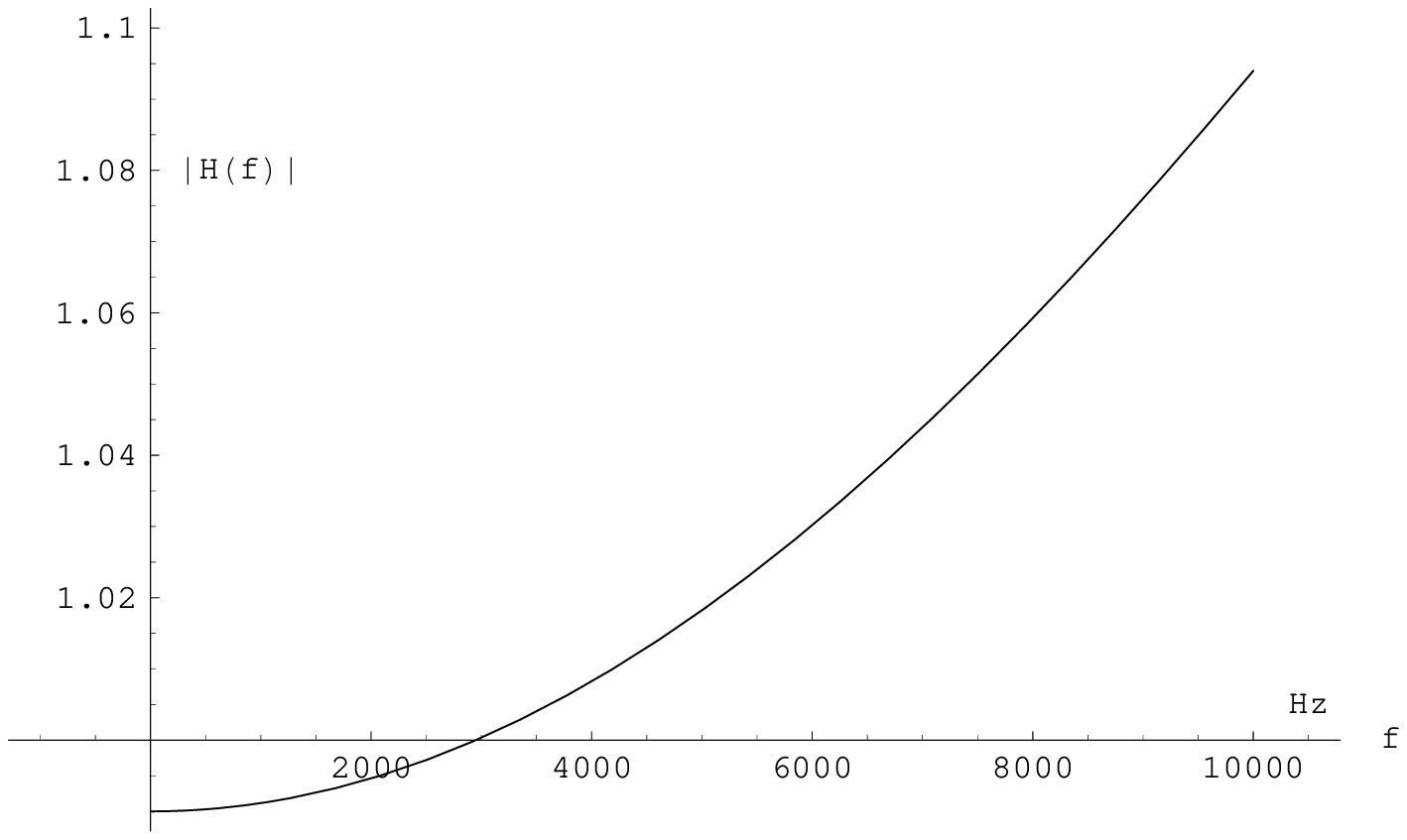}

\caption{the absolute value of the longitudinal response function (\ref{eq: segnale totale lungo z})
of the LIGO interferometer ($L=4Km$) to a GW arising from the \textbf{$R^{2}$}
high order gravity theory and propagating with a speed of $0.1c$
(non relativistic case). }
\end{figure}
\begin{figure}
\includegraphics{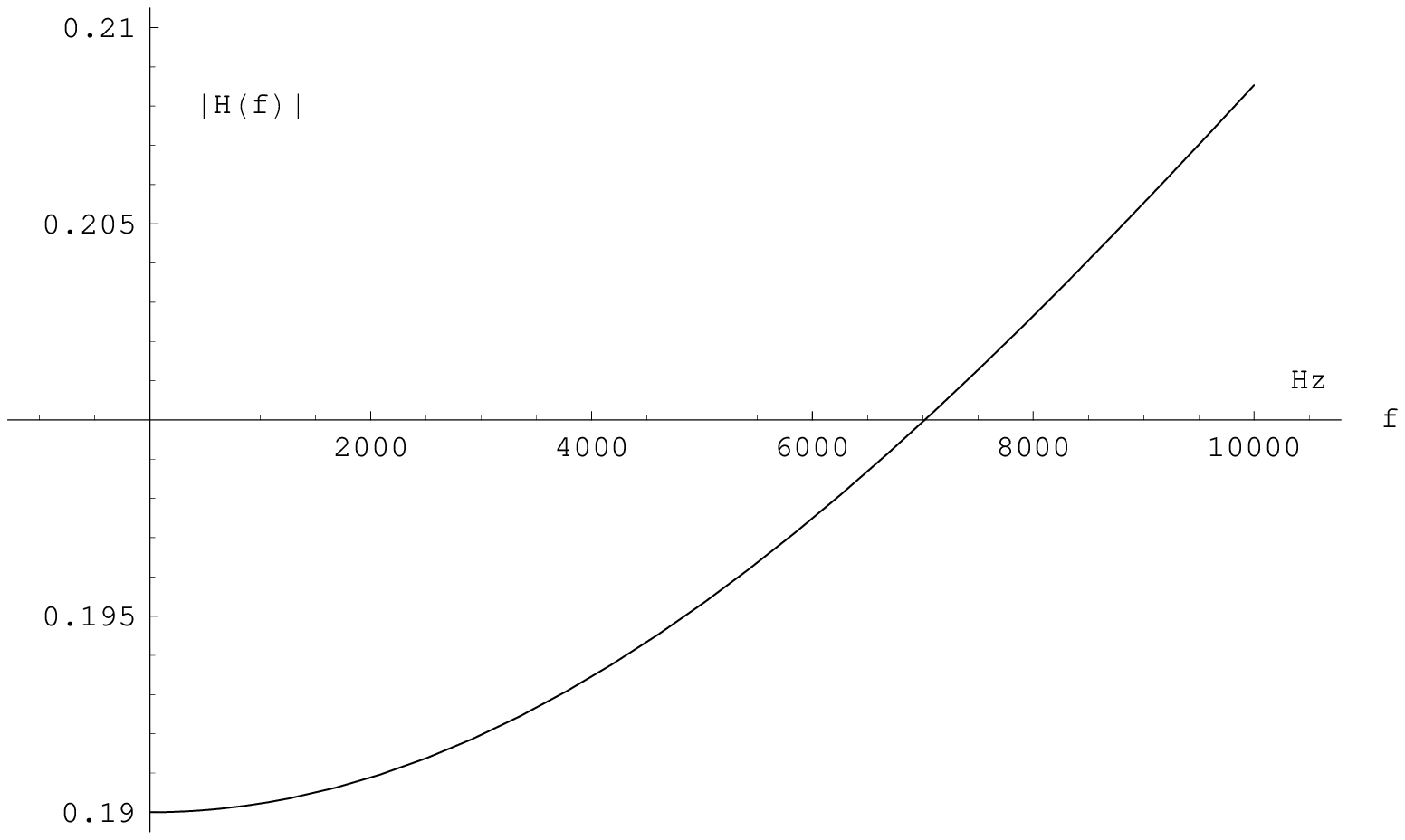}

\caption{the absolute value of the longitudinal response function (\ref{eq: segnale totale lungo z})
of the LIGO interferometer ($L=4Km$) to a GW arising from the \textbf{$R^{2}$}
high order gravity theory and propagating with a speed of $0.9c$
(relativistic case). }
\end{figure}
\begin{figure}
\includegraphics{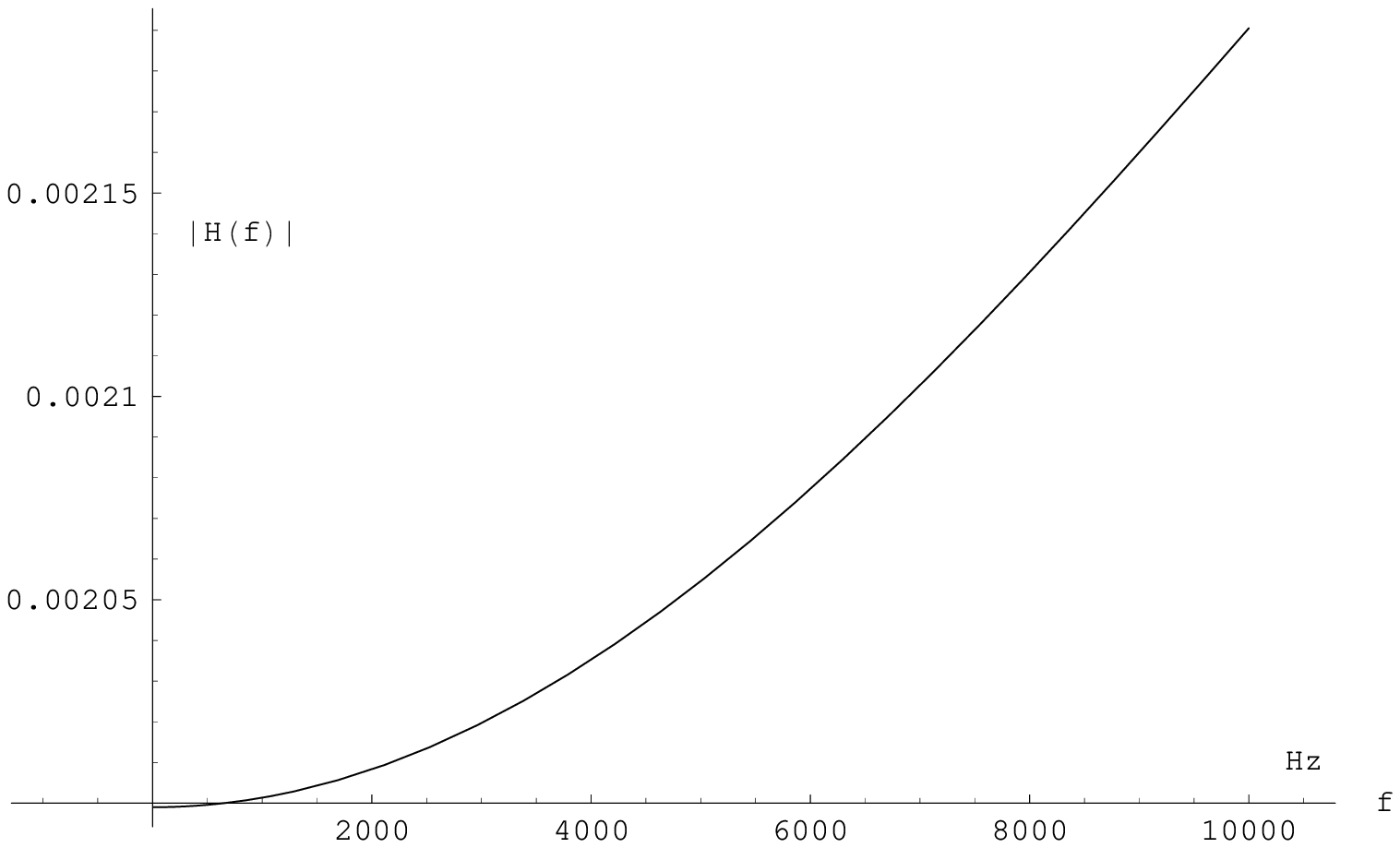}

\caption{the absolute value of the longitudinal response function (\ref{eq: segnale totale lungo z})
of the LIGO interferometer ($L=4Km$) to a GW arising from the \textbf{$R^{2}$}
high order gravity theory and propagating with a speed of $0.999c$
(ultra relativistic case). }
\end{figure}

\section{Conclusions}

We have shown that from the \textbf{$R^{2}$} high order gravity theory
it is possible to produce, in the linearized approch, particles which
can be seen like massive modes of gravitational waves. The presence
of the mass generates a longitudinal force in addition of the transverse
one which is proper of the massless gravitational waves and the response
an interferometer to the effect has been computed. The presence of
the mass could also have important applications in cosmology because
the fact that gravitational waves can have mass could give a contribution
to the dark matter of the Universe. As a final remark,  we recall
that the potential detection of a longitudinal component of GWs could
be, in principle, an useful tool to discriminate among several gravity
theories which are today considered.

\section{Acknowledgements}

I would like to thank Salvatore Capozziello, Maria Felicia De Laurentis
and Franceso Rubanu for helpful advices during our work. I have to
thank the European Gravitational Observatory (EGO) consortium for
the using of computing facilities.

\end{document}